# Thermodynamics with thermodynamic variable first-passage time. II. Dynamics of explosive boiling of superheated liquid, critical indices, fractional calculus.

V. V. Ryazanov ORCID 0000-0002-5308-3212
Institute for Nuclear Research, pr. Nauki, 47 Kiev, Ukraine

Within the framework of stochastic first-passage time thermodynamics (TFPT) and power-law distributions (Mittag-Leffler and Levy), the appearance of fractional Caputo derivatives in hydrodynamic equations is substantiated. The developed framework is applied to the problem of explosive boiling of a superheated liquid: an integral formula for nonlocal entropy production is derived and the critical index of dissipation intensity is calculated, along with qualitative changes in these parameters upon introducing external pressure. It is found that dissipation at a given instant is strictly determined by the integral history of the macroscopic matter flow J(t) over the entire evolutionary interval up to the moment of phase explosion. The calculated critical dissipation index z strictly describes the kinetic catastrophe and the violation of Prigogine's minimum entropy production principle near spinodals.

## 1. Introduction

In this paper, an example of the application of the general theory presented in Ref. [1] is given. In Refs. [2–6], distributions containing the first-passage time (FPT) $T_{fpt}$ (lifetime) were introduced. In Refs. [7–9], based on this kind of distributions, thermodynamics with the first-passage time Ref. [9] was proposed. In Ref. [1], this approach, as in Ref. [9], is combined with Zubarev's nonequilibrium statistical operator (NSO) method and extended nonequilibrium thermodynamics (EIT) Ref. [10]. The unity of these three concepts creates a closed scheme in which the form of the first-passage time distribution $\rho(t_{fpt})$ corresponds to microscopic fluctuations and the geometry of the system. From it, a generalized thermodynamic potential $\Phi(\gamma)$ is constructed, yielding the macrovariable of the first-passage time $T_{fpt}$. Through the closure procedure of Zubarev's NSO [12-14], this variable determines the specific mathematical form of the memory kernels K(t). Closed non-Markovian Maxwell-Cattaneo equations with these memory kernels describe real, physically observable energy transfer and phase transitions in complex media.

In Refs. [9] and [1] a synthesis of extended nonequilibrium thermodynamics and Zubarev's NSO method is carried out in the first-passage time formalism of Refs. [7-9]: from stochastic trajectories to nonlinear transport equations.

In Ref. [1], as in Ref. [10], the limitations of Onsager's classical thermodynamics in describing high-intensity, fast-moving processes and low-dimensional systems (nanochannels, polymer matrices) are noted, as well as the need to strictly take into account memory effects (non-Markovian properties) and spatial non-locality. Also important are the microscopic justification of the phenomenological parameters, the non-closed nature of the macroscopic balance equations (the problem of higher moments), and other important points; they are addressed in Ref. [1] and in this paper.

The theoretical basis and the generalized Gibbs ensemble in articles [2-6] are based on the definition of the first transit time (FPT) as a fundamental stochastic variable characterizing the "lifetime" of a local state. It should be noted that the random variable $T_{fpt}$ in the nonequilibrium case has the same generality as the energy in the equilibrium case, since the kinetic equations for the distributions of energy and $T_{fpt}$ are conjugate. The mathematical formalism for constructing the generalized partition function is defined through the Laplace transform of the FPT distribution density $T_{fpt}$, the partition function $Z_{fpt}(\beta,\gamma)$. The thermodynamic force $\gamma$ in this case takes on the physical meaning of a fluctuation regulator and a relaxation intensity parameter.

The introduction of the generalized thermodynamic potential (free energy FPT) $\Phi(\beta,\gamma) = -\ln Z_{fpt}$ and the derivation of the conjugate macroscopic variables by differentiation correspond to Gibbs thermodynamics and the relations of Ref. [11].

The procedure for closing transport equations in Zubarev's NSO method involves a generalized description. The construction of a quasi-equilibrium (relevant) [14] statistical operator $\rho_{rel}(t)$ in the proposed version of nonequilibrium thermodynamics is generalized by explicitly including the first-passage time value $T_{fpt}$.

The Liouville equation and the derivation of macroscopic balance equations for flows lead to a solution to the non-closure problem. Thermodynamic consistency conditions and the expression of the force γ in terms of the measured mean $\langle T_{fpt} \rangle$ are used, as is done in the NSO method.

The modification of the non-Markovian transport equations in the Maxwell–Cattaneo equation carried out in Refs. [9] and [1], together with a comparison with EIT, yields a rigorous microscopic derivation of this equation with the replacement of the constant phenomenological relaxation time τ by the state functional $\langle T_{fpt} \rangle$. In Ref. [1], the nature of the emerging macroscopic nonlinearity is considered.

In Ref. [1] the analytical connection between the Laplace image of the memory core $\tilde{K}(s)$ and the potential $\Phi(s)$ is also deduced, which, within the framework of renewal theory, leads to the mathematical structure of the memory cores.

The transition in dissipation regimes from the Markov limit (Debye delta-shaped kernel) to non-Markov fractional-power decay of correlations $K(t) \propto t^{-(2-\alpha)}$ is discussed.

Using renewal theory, we can establish that the Laplace image of the memory kernel $\tilde{K}(s)$ in generalized kinetic equations is uniquely determined by the structure of the generalized thermodynamic potential $\Phi(s)$, which corresponds to the causal justification of memory kernels. This eliminates the need to introduce phenomenological approximations for memory kernels. The thermodynamics of lifetime fluctuations treats the first-passage time as a fully-fledged macroscopic variable, which distinguishes it from standard kinetic approaches. Using second derivatives of the generalized potential $\partial^2\Phi / \partial\gamma^2$, this formalism allows us to calculate the variance, as well as higher fluctuation moments of the system's lifetime, which is critical at the nanoscale when describing viscoelasticity and anomalous diffusion.

In this paper, we consider the theory of critical slowing down of metastable phases based on the results obtained in Ref. [1]. Near phase transition points and spinodals, where the Markov approximation completely breaks down and FPT distributions exhibit power-law "heavy tails," the parameter γ acts as a natural thermodynamic regulator. Scaling analysis of the potential $\Phi(\epsilon,\gamma)$ allows us to determine the dynamic universality classes of systems through critical indices of the metastable phase lifetime.

The entropy production of stochastic transitions behaves as follows. Local entropy production of a metastable system is strictly linked to the dynamics of compression of the accessible phase space when approaching the stochastic boundary: $\sigma = k_B \gamma d\langle T_{fpt} \rangle / dt$ (8) Ref. [1]. This relation links macroscopic energy dissipation (e.g., during explosive boiling or crystallization) with the microscopic search for the critical size of the nucleus.

Using heavy-tailed power-law distributions for FPT leads to a transition to the fractional Caputo derivative. Substituting the Laplace transform of the Mittag-Leffler distribution into the memory kernel structure corresponds to the derivation of a fractional differential equation of hydrodynamic transport of order $(\alpha\ )$ $(0 < \alpha < 1)$.

These equations describe intermittency and fractal memory. The physical meaning of the fractional operator $\mathbf{D}_t^{\alpha}$ is that it acts as an integrator of the history of the viscous and thermal resistance of the medium.

The article is organized as follows. Section 2 examines the boiling of superheated liquids and nonlinear dissipation. The critical indices of these processes are discussed in Section 3. Section 4 compares heavy-tailed distributions for liquid boiling. Section 5 discusses the consequences of the heavy-tailed distributions used, which lead to fractional kinetic equations, nonlocality, and singularity of dissipation during boiling. Fractional calculus in the physics of metastable media with memory is considered. Section 6 examines the dynamics of explosive boiling of superheated liquids in more detail than Section 2. Section 7 discusses the versatility and generality of the proposed approach. Section 8 concludes.

## 2. An example of boiling of superheated liquid and nonlinear dissipation

In Ref. [1], an example of explosive boiling of a superheated liquid was mentioned. To illustrate the capabilities of the approach of Refs. [9] and [1], taking into account Zubarev's NSO method, we will consider the process of boiling of a superheated liquid (a first-order phase transition). In the classical nucleation theory (Frenkel-Zel'dovich) Ref. [15], boiling is described as the overcoming of the critical size $R_{cr}$ of the Gibbs potential barrier by a free nucleus, but this theory poorly takes into account strong fluctuations in small volumes and the memory effects of the medium.

The method based on the thermodynamics of the first-passage time (TFPT) Refs. [9, 1] and the nonequilibrium statistical operator (NSO) Refs. [12–14] describes the boiling of a superheated liquid in a fundamentally different way—through the dynamics of the random "lifetime" of a metastable state. The connection between the approach of Refs. [9, 1] and the NSO method was carried out in Ref. [16].

The practical application of the synthesized non-Markovian formalism to the description of first-order phase transitions allows for a fundamentally different view of the kinetics of nucleation in metastable media. Within the framework of the classical Frenkel-Zel'dovich nucleation theory Ref. [15], the boiling of a superheated liquid is interpreted as the diffusional overcoming of the work of formation of a critical bubble $W_{cr} = W\left(R_{cr}\right)$ by a fluctuating nucleus of radius R. This barrier is determined by the classical Gibbs potential:

$$W(R) = 4\pi R^2 \sigma_{sur} - \frac{4}{3}\pi R^3 \Delta P , \tag{1}$$

where $\sigma_{sur}$ is the surface tension, and $\Delta P$ is the pressure difference between the phases. In the stochastic formalism under consideration, the critical radius point $R = R_{cr} = 2\sigma_{sur} / \Delta P$ defines the position of the absorbing hypersurface $\partial\Omega$ in the configuration space Ref. [1]. The random lifetime of the metastable phase $T_{fpt}$ is the time of first reaching this boundary. Taking into account the thermal inertia and viscous resistance of the medium, the relaxation of fluctuations near the barrier obeys the non-exponential Mittag-Leffler law (28), which for large values of t decreases as a power law $t^{-1-\alpha}$. In terms of the NSO method, this transforms the Zeldovich Markov equation into a non-Markov transport equation with a fractional Caputo derivative of order $\alpha$ $(0 < \alpha \leq 1)$:

$$\mathbf{D}_t^{\alpha} f(R,t) = \frac{1}{\tau_{cr}^{\alpha}} \frac{\partial}{\partial R}\left[ D_R \frac{\partial f}{\partial R} + \frac{D_R}{k_B T} \frac{\partial W(R)}{\partial R} f \right], \tag{2}$$

where $\tau_{cr} \propto \exp(W_{cr} / k_B T)$ is the characteristic Kramers waiting time, and the fractional derivative operator Caputo $\mathbf{D}_t^{\alpha}$ of order $\alpha$, defined as: $\mathbf{D}_t^{\alpha} J(t) = \frac{1}{\Gamma(1-\alpha)} \int_0^t \frac{1}{(t-t')^{\alpha}} \frac{\partial J(t')}{\partial t'} dt'$ (32), performs integration over the entire history of the viscoelastic resistance of the fluid.

The central result of combining the equations for the macroscopic flow of matter $J(t) = -D \cdot (\partial \langle T_{fpt} \rangle / \partial x)$ and the stochastic entropy production $\sigma = k_B \gamma \cdot (d\langle T_{fpt} \rangle / dt)$ (8) Ref. [1] is the derivation of the final nonlinear equation for dissipation during boiling. Eliminating the conjugate force γ using the thermodynamic matching procedure Ref. [1], we obtain a rigorous expression for the local entropy production, expressed directly in terms of the integral characteristics of the flow (44), see derivation (44):

$$\sigma(t) = \frac{k_B \cdot (\alpha \tau_{cr}^{\alpha})^{\frac{1}{1-\alpha}}}{nD} \cdot \frac{J^2(t)}{\left[\int_t^{t_{vzb}} J^2(t')dt'\right]^{\frac{1}{1-\alpha}}}, \tag{3}$$

where $t_{vzb}$ is the moment of complete phase decay (explosion). This formula demonstrates the profound temporal nonlocality of dissipative processes in a metastable medium: entropy generation at the current moment t depends nonlinearly on the flow intensity over the entire future interval up to the moment of boiling.

As the spinodal is approached, when $t \to t_{vzb}$, the denominator of the equation tends to zero, causing a singular power-law burst of dissipation $\sigma(t) \to \infty$, the scale of which is determined by the critical index of the boiling intensity $z = (\alpha^2 - \alpha + 1)/(1-\alpha)$ (derivation of z in Section 3). The introduction of external pressure $P_{ext}$ (Section 5) modifies the phase space geometry, shifting the boundary coordinate $R_{cr}(P_{ext})$ to the right and causing an exponential increase in the barrier waiting time $\tau_{cr}(P_{ext})$. Concurrently, increasing pressure reduces the non-Markovian parameter $\alpha \to 0$, which leads to an increase in the denominator of the critical index $z(P_{ext})$ and, consequently, to a smoothing of the singularity of the dissipative burst. Thus, external pressure transforms the explosive nature of homogeneous nucleation into a regime of smooth, kinetically controlled phase transformation.

Thus, microscopic parameters (nucleus radius, barrier) are fully linked with macroscopic nonlinear dissipation and the influence of pressure.

Formulation of the problem in terms of the TFPT. The superheated liquid is at a metastable free energy minimum. It must overcome the activation barrier (the work of formation of a critical nucleus, $W_{cr}$). In phase space, the process coordinate is the radius of the vapor nucleus, R. The absorption boundary is the point $R=R_{cr}$, which is the absorption boundary. As soon as the density fluctuation generates a bubble of radius $R_{cr}$, it begins to grow uncontrollably—the liquid boils explosively. The random variable is the first-passage time, $T_{fpt}$—this is the random time it takes for the nucleus, fluctuating under the impact of liquid molecules, to first reach the critical size, $R_{cr}$.

How is Zubarev's NSO method applied? Traditionally, boiling kinetics is described by a Fokker-Planck equation, where the diffusion coefficients of the nuclei by size are assumed to be constant. The Refs. [1, 9] method constructs a generalized relaxation operator, introducing the parameter γ, associated with the liquid's superheated lifetime:

$$\rho_{rel} = \exp\left(-\Phi - \beta\mathcal{H} - \gamma T_{fpt}(R)\right). \tag{4}$$

The parameter γ here physically determines the intensity of the thermal "push" of the liquid toward phase transition. The more the liquid is superheated relative to its boiling point, the greater the value of the thermodynamic force γ.

The application of Zubarev's combined NSO method and first-passage time thermodynamics (TFPT) to systems near phase transitions (including decays of metastable states) is one of the most powerful aspects of the theory Refs. [1, 9]. In these regions, classical thermodynamics and standard kinetic approaches fail, as lifetime fluctuations become commensurate with the lifetime itself.

The Refs. [1, 9] method describes this area through three key steps.

1. Transition from an exponential distribution to "heavy-tailed" ones. Far from the transition points, the distribution of the first-passage time (FPT) of a metastable state is a typical exponential one $\rho(T_{fpt}) \propto e^{-t/\tau}$, where τ is a fixed relaxation time (e.g., the Kramers time).

Near phase transition points (critical points, spinodal lines), fluctuations in the order parameter increase sharply. This is known as critical slowing down. The system becomes "stuck" in a metastable state for a long time and can then instantly collapse into a stable phase. In the Refs. [1, 9] method, this is reflected by a change in the mathematical form of the partition function: the FPT distribution acquires a power-law form with a "heavy tail" (Lévy, Pareto, or Weibull distributions):

$$\rho(t_{fpt}) \propto t_{fpt}^{-(1+\alpha)}, \quad 0 < \alpha < 1. \tag{5}$$

For such distributions, the classical mean time $\langle T_{fpt} \rangle$ or its variance formally tends to infinity.

2. Thermodynamic regulator γ as a control parameter. At this point, standard kinetic equations become unsolvable. However, in Zubarev's modified NSO, the parameter γ (the thermodynamic force associated with the FPT) acts as a natural mathematical regulator.

Since the generalized statistical sum (in a system with fixed energy) is constructed as the Laplace transform with respect to the parameter γ:

$$Z_{fpt}(\gamma) = \int_0^\infty \rho(t_{fpt}) e^{-\gamma t_{fpt}} dt_{fpt}, \tag{6}$$

the multiplier $e^{-\gamma t_{fpt}}$ effectively "cuts off" the infinite tails of the fluctuation distributions at the critical point.

Near a phase transition, the parameter γ physically describes the magnitude of the thermodynamic stimulus (pressure) that the external environment exerts on the metastable phase to force it to transition to the stable phase. The value $\gamma \to 0$ corresponds to a strict critical state (where fluctuations are infinite), and finite values of γ allow us to calculate the thermodynamic potentials of the metastable phase even under conditions of gigantic fluctuations.

3. Calculation of entropy production singularities. In classical metastable systems, entropy production σ is assumed to be constant or smoothly changing. The method of Refs. [1, 9] relates lifetime fluctuations to entropy production directly through the second derivatives of the potential $\Phi(\gamma)$:

$$\langle (\Delta T_{fpt})^2 \rangle = -\frac{\partial^2 \Phi}{\partial \gamma^2}. \tag{7}$$

At the phase transition point, fluctuations $\langle (\Delta T_{fpt})^2 \rangle$ undergo a jump or diverge according to a power law (characterized by critical indices). The entropy balance relation is used:

$$\sigma = k_B \gamma \frac{d\langle T_{fpt} \rangle}{dt}. \tag{8}$$

This means that entropy production in a metastable system near a phase transition becomes a non-stationary, highly fluctuating quantity. Zubarev's NSO method allows us to strictly link fluctuations in this entropy production to the nucleation of a new phase.

Abstract: Instead of ignoring metastability or describing it approximately, Zubarev's method and the results of Refs. [1, 9] consider the lifetime of the metastable state itself as a fluctuating thermodynamic variable. This allows one to calculate the phase stability boundaries (spinodals and binodals) taking into account the spectrum of fluctuations, not just their average values.

How does the method describe critical boiling anomalies?

A. Accounting for the "heavy tails" of the distribution (superfluctuations). If the superheat is low, bubbles nucleate rarely, the $T_{fpt}$ distribution is exponential, and the average boiling time is enormous. However, as the spinodal limit (the absolute instability of the liquid) is approached, a barrier $W_{cr} \to 0$ (on the spinodal line, at the moment the liquid reaches the boundary of thermodynamic catastrophe $W_{cr} = 0$) appears.

Lifetime fluctuations become gigantic and acquire a power-law character (Lévy distribution): individual regions of the liquid may boil instantly, while others may maintain metastability for an abnormally long time. The generalized partition function $Z_{fpt}(\gamma)$, using the γ regulator, collects information about all these trajectories, preventing mathematical divergence of the equations.

B. Calculating non-Markovian properties and memory effects. The boiling of a superheated liquid is not an instantaneous event. The growth of a vapor bubble requires local heat removal (the phase transition absorbs the latent heat of vaporization) and deformation of the surrounding liquid (viscous drag). This gives rise to memory effects: the liquid "remembers" how the bubble grew an instant ago.

Within the framework of the Refs. [1, 9] method, this introduces a dependence of the parameter γ on the process history via a frequency-dependent memory kernel K(ω). As a result, the equation for the average boiling time becomes an integro-differential equation. The method demonstrates that the thermal inertia of the surrounding liquid significantly delays the actual boiling time compared to classical Markov estimates.

B. Connection with entropy production. At the moment of formation of a critical bubble, local entropy production increases sharply: $\sigma = k_B \gamma d\langle T_{fpt}\rangle / dt$ (8).

Since near the spinodal, the quantity $\langle T_{fpt}\rangle$ depends nonlinearly on density fluctuations, entropy production experiences a sharp spike (singularity). Zubarev's method and Refs. [1, 9] allow one to rigorously calculate this spike, linking the macroscopic energy dissipation during explosive boiling with the microscopic fluctuations of molecules. A more detailed discussion of these issues is provided in Section 6.

An alternative example: polymer crystallization. If we apply the same method to polymer crystallization, the specificity will be in the physical sense of the barrier. Polymer chains are long and tangled, requiring time for segmental rearrangement (relaxation) to fit into the crystal lattice.

Here, the memory effects (non-Markovian properties) inherent in the Φ(γ) potential play an even more fundamental role than in liquids. The first-passage time $T_{fpt}$ describes the time required for a macromolecule segment to unravel and attach to the surface of the growing crystal. The γ parameter in this case is directly related to the thermodynamic rigidity of the chain and the cooling rate of the polymer.

## 3. Calculation of critical indices of the lifetime of the metastable phase

The calculation of critical indices of the lifetime of a metastable phase near phase transition points (or spinodals) through a generalized thermodynamic potential is based on the apparatus of Large Deviations Theory and the theory of scale invariant (scaling) analysis Refs. [17-19].

In the Refs. [1, 9] method, this is realized through the study of the mathematical properties of the generalized Mathieu potential as the control parameters of the system tend to critical values. Below is the step-by-step mathematical logic for calculating these indices.

Step 1: Scaling hypothesis for the generalized potential. Let be $\epsilon = (T - T_c)/T_c$ the dimensionless distance to the critical point of the phase transition (e.g., to the critical boiling temperature or spinodal) Ref. [18].

By analogy with Landau's equilibrium theory—fluctuation theory—we use the similarity (scaling) hypothesis for the dynamic potential $\Phi(\epsilon,\gamma)$. Near the critical point, the potential splits into a regular part and a singular (anomalous) part:

$$\Phi(\epsilon,\gamma) = \Phi_{reg} + \Phi_{sing}(\epsilon,\gamma). \quad (9)$$

The singular part in (9), due to scale invariance near the transition, is a homogeneous function of its arguments:

$$\Phi_{sing}(\lambda^a \epsilon, \lambda^b \gamma) = \lambda \Phi_{sing}(\epsilon,\gamma), \quad (10)$$

where *a* and *b* are critical exponents (dynamic scaling factors) Ref. [19].

By choosing a scaling factor $\lambda = \epsilon^{-1/a}$, we can rewrite the potential in standard self-similar form:

$$\Phi_{sing}(\epsilon,\gamma) = \epsilon^{1/a} f\left(\frac{\gamma}{\epsilon^{b/a}}\right) = \epsilon^{2-\alpha_{dyn}} f\left(\frac{\gamma}{\epsilon^{\Delta}}\right), \quad (11)$$

where $\alpha_{dyn}$ is an analogue of the critical heat capacity index, responsible for the potential anomaly, $\Delta = b/a$ the so-called “crossover” or dynamic exponential index.

Step 2: Relationship with the critical lifetime index. The lifetime of a metastable state (macroscopic mean first-passage time $\langle T_{fpt}\rangle$) is found as the first derivative of the potential with respect to the conjugate force $\gamma$:

$$\langle T_{fpt}\rangle = \frac{\partial\Phi(\epsilon,\gamma)}{\partial\gamma}. \tag{12}$$

Differentiating our scaling relation (11) with respect to $\gamma$, we obtain:

$$\langle T_{fpt}\rangle = \epsilon^{2-\alpha_{dyn}-\Delta} f'\left(\frac{\gamma}{\epsilon^{\Delta}}\right). \tag{13}$$

If we are looking for the free lifetime of the system (without external critical "pressure", i.e. at $\gamma \to 0$), then function $f'(0)$ from (13) turns into a constant, and we obtain a pure power-law dependence of the lifetime on the proximity to the critical point:

$$\langle T_{fpt}\rangle \propto \epsilon^{-\tau_{fpt}}. \tag{14}$$

From here we find the critical index of the lifetime of the metastable phase $\tau_{fpt}$ through the indices of the potential:

$$\tau_{fpt} = \Delta + \alpha_{dyn} - 2. \tag{15}$$

Physically, the index $\tau_{fpt}$ describes the critical slowing down of a system. The closer we get to the phase transition $\epsilon \to 0$, the longer and more intensely the metastable state fluctuates before decaying.

Step 3: Calculating the lifetime fluctuation index. The most important advantage of the Refs. [1, 9] method is the ability to calculate the lifetime fluctuation index (variance), which is simply lost in classical thermodynamics. The second derivative of the potential yields the square of the fluctuations:

$$\left\langle (\Delta T_{fpt})^2 \right\rangle = \frac{\partial^2\Phi}{\partial\gamma^2} = \epsilon^{2-\alpha_{dyn}-2\Delta} f''\left(\frac{\gamma}{\epsilon^{\Delta}}\right). \tag{16}$$

When $\gamma \to 0$ fluctuating, they diverge from their critical index $\mu$:

$$\langle (\Delta T_{fpt})^2 \rangle \propto \epsilon^{-\mu}, \tag{17}$$

where $\mu = \mathbf{2}\Delta + \alpha_{\mathbf{dyn}} \mathbf{-2}$. Comparing the two indices, we see the universal scaling relation:

$$\mu = \tau_{fpt} + \Delta. \tag{18}$$

The critical boiling intensity index z is obtained mathematically rigorously by comparing the asymptotic behavior of the macroscopic flow J(t) and the thermodynamic force γ as the phase explosion approaches. The calculation relies on large deviation theory, dynamic scaling methods, and the fractal time structure of the Mittag-Leffler distribution.

Below is a step-by-step analytical procedure for calculating the index z.

Step 1. Time scaling near the explosion point. Let be $\Delta t = t_{vzb} - t$ the time remaining until complete phase decay (boiling). In the critical region (at $\Delta t \to 0$), the average remaining lifetime of the metastable state $\langle T_{fpt}\rangle_t$ decays according to a power law dictated by the fractal non-Markovian exponent of the medium $\alpha$ $(0 < \alpha < 1)$:

$$\langle T_{fpt}\rangle_t \propto (\Delta t)^{\alpha}.$$

Differentiating this expression with respect to physical time t (taking into account that $d(\Delta t)/dt = -1$), we find the rate of change of lifetime:

$$\frac{d\langle T_{fpt}\rangle_t}{dt} \propto -(\Delta t)^{\alpha-1}.$$

Step 2. Finding the asymptotic behavior of the thermodynamic force γ. Below, in expressions (37)-(38), using the Laplace transform for the Mittag-Leffler distribution, we derive the equation of state relating the conjugate force γ and the mean first-passage time: $\langle T_{fpt}\rangle \approx \alpha\tau_{cr}^{\alpha}\gamma^{\alpha-1}$. From here, we express the force γ through the current macroscopic lifetime (39): $\gamma \propto \langle T_{fpt}\rangle^{-\frac{1}{1-\alpha}}$. Substituting the time scaling $\langle T_{fpt}\rangle \propto (\Delta t)^{\alpha}$ from Step 1 into this equation, we obtain how the force γ depends on the temporal distance to the explosion $\Delta t$:

$$\gamma(t) \propto \left((\Delta t)^{\alpha}\right)^{-\frac{1}{1-\alpha}} = (\Delta t)^{-\frac{\alpha}{1-\alpha}}.$$

Step 3. Calculation of the divergence of entropy production $\sigma(t)$. The basic relation of the entropy balance in the formalism of Zubarev and Refs. [1, 9] has the form (8): $\sigma(t) = k_B\gamma(t)\dfrac{d\langle T_{fpt}\rangle_t}{dt}$. Let's substitute into this formula the obtained asymptotics for the force $\gamma(t)$ (from Step 2) and for the derivative $d\langle T_{fpt}\rangle / dt$ (from Step 1):

$$\sigma(t) \propto (\Delta t)^{-\frac{\alpha}{1-\alpha}} \cdot (\Delta t)^{\alpha-1}.$$

Thus, the law of divergence of entropy production at the pre-explosive moment takes on a pure power form:

$$\sigma(t) \propto (\Delta t)^{-z},$$

where the exponent is the desired critical index of boiling intensity z:

$$z = \frac{\alpha^2 - \alpha + 1}{1-\alpha}.$$

Physical analysis of the obtained z-index:

- The degree of catastrophicity of the process: since for media with memory 0<α<1, the denominator 1-α is always positive and less than one. This makes the index z significantly greater than one (for example, at α=0.5, the index z=1.5). Such a high degree of divergence means an avalanche-like, explosive release of entropy immediately before the phase transition.
- Markov limit (α→1): if the system loses memory and becomes classical Markov (α→1), the denominator tends to zero, and the index $z \to \infty$. Physically, this means an instantaneous, singular break (a mathematical singularity of the delta function type). The fractional parameter α<1 “smears” this catastrophe in time, making it accessible for continuous macroscopic description.

What is the practical physical meaning of this calculation? Definition of the universality class. Just as in equilibrium thermodynamics, indices $\alpha, \beta, \gamma, \delta$ determine the universality class of a system (e.g., the Ising or Heisenberg models), dynamic indices $\tau_{fpt}$, $\mu$ allow us to group completely different processes (boiling of liquids, crystallization of polymers, or magnetization reversal of ferromagnets) into unified classes of dynamic behavior Ref. [19].

Prediction of the catastrophe point (critical limit). Knowing experimentally or from molecular dynamics how the fluctuations of the lifetime of the metastable phase change, and knowing the calculated index $\mu$, one can predict with high accuracy the spinodal point (the boundary of absolute instability) when an instantaneous explosive phase transition (e.g., instantaneous boiling) will occur Ref. [20].

## 4. Comparison of the use of distributions with "heavy tails" to describe the boiling of liquids

Macroscopic transport equations are derived from the microscopic Liouville equation through a statistical averaging procedure. This is a rigorous mathematical derivation of the laws of hydrodynamics and thermodynamics from the laws of motion of individual molecules. The Liouville equation describes the precise, yet unobservable, dynamics of trillions of particles. Zubarev's method allows us to transform it into understandable equations for macroscopic flows (heat, mass, momentum).

How does Zubarev derive equations from the Liouville equation? The process consists of four sequential steps: a) defining microscopic variables: the exact coordinates and momenta of all molecules are specified. b) applying the Liouville operator: microscopic time derivatives of energy or particle densities are calculated from the Liouville equation. c) constructing the nonequilibrium statistical operator: the Liouville equation is solved with a special boundary condition (the causality principle). This is how Zubarev finds the nonequilibrium statistical operator (NSO). In d) statistical averaging, the microscopic flow equations are averaged using this NSO.

In NSO, as shown in Ref. [16], there is an averaging over the distribution of the system's lifetime. The exponential and power-law distributions for the lifetime were considered above. The gamma distribution is the product of such factors. The gamma distribution can be considered as an intermediate, transitional regime between a pure exponential distribution and anomalous power-law distributions with "heavy tails." Moreover, in the physics of metastable systems and the theory of random processes, this class of distributions plays a fundamental role.

The density of the gamma distribution has the form:

$$\rho(t) \propto t^{\alpha-1} e^{-t/\tau}. \tag{19}$$

It is a direct product of the power-law factor $t^{\alpha-1}$ (5) and exponential smoothing $e^{-t/\tau}$. Its intermediate nature manifests itself depending on the observation time scales: on small and medium times $t \ll \tau$. The exponential term $e^{-t/\tau} \approx 1$, and the distribution behaves as a purely power-law $\rho(t) \propto t^{\alpha-1}$. The system exhibits strong fluctuations characteristic of non-Markovian processes. On large times, as $t \gg \tau$, the exponential tail inevitably suppresses the power-law factor. The distribution transitions to the exponential regime. This means that all higher moments (mean time, variance) remain finite, and the "infinite" fluctuations of the critical point are effectively "cut off" on the scale $\tau$.

In the context of the Refs. [7-9] method, the gamma distribution ideally describes the situation when the system is close to a phase transition, but has not yet reached it. The scale $\tau$ in this case is directly related to the magnitude of the thermodynamic force γ.

Physical analogue: Truncated Lévy Flight distribution. From a rigorous perspective, the gamma distribution isn't the only candidate for this role. The most accurate and physically sound "intermediate" variant between the exponential and pure power (Lévy) is the truncated Lévy Flight distribution (or exponentially truncated stable law). Its Laplace transform (which in TFPT is identical to the generalized partition function $Z_{fpt}(\gamma)$) has the structure:

$$\rho(s) = \exp\left(-A[(s+\gamma_0)^{\alpha} - \gamma_0^{\alpha}]\right). \tag{20}$$

How it works as a bridge. Pure power-law regime (phase transition point/spinodal). At $\gamma_0 \to 0$ in (20), we obtain a pure Levy law $\tilde{\rho}(s) \propto \exp(-As^{\alpha})$. Temporal correlations decay slowly (as a power law), and the lifetime fluctuations of the metastable phase diverge. Exponential regime (far from the transition). At very large values $\gamma_0$ (strong external field or severe overheating), the Laplace transform expansion (20) becomes linear in s, which returns us to the classical Debye exponent and the Markov process. Intermediate regime (subcritical state). At moderate values $\gamma_0$, the distribution (20) looks like a fractal power-law process on short times, and on giant times, it smoothly decays exponentially.

Another important candidate: the Mittag-Leffler distribution. In non-Markovian thermodynamics and continuous random walk (CTRW) theory, when memory effects are introduced, another function often emerges that acts as an intermediate bridge: the Mittag-Leffler distribution (28).

Its distribution density behaves exactly the opposite (compared to the gamma distribution). At short times $t \to 0$, it decays exponentially (tends to classical relaxation). At long times $t \to \infty$, it transforms into a pure power law $\rho(t) \propto t^{-(1+\alpha)}$.

This distribution describes metastable systems with deep memory "traps": at first the process proceeds normally, but if the system gets stuck in a local minimum, the exit latency acquires a heavy power-law tail.

Considering the gamma distribution (or its generalization in the form of a truncated Levy distribution) as an intermediate stage is absolutely correct. Within the framework of the formalism of Refs. [1, 7-9], the transition of a system from an exponential distribution to a power-law distribution through gamma-like structures mathematically describes the trajectory of the system's movement to a critical point (the point of catastrophe).

In the framework of first-passage time thermodynamics (TFPT), the scale τ and the thermodynamic force γ are related by an inverse proportionality relation:

$$\tau \propto \frac{1}{\gamma}. \tag{21}$$

This relationship follows from the very structure of the generalized distribution Refs. [1, 9] and the properties of the Laplace transform. Below is a physical and mathematical explanation.

Mathematical comparison of factors. The gamma distribution density describing the transition mode of the metastable phase lifetime has the form (19): $\rho(t_{fpt}) = C \cdot t_{fpt}^{\alpha-1} e^{-\frac{t_{fpt}}{\tau}}$. On the other hand, in Zubarev's nonequilibrium statistical operator (NSO) method, the generalized probability density for the first-passage time contains an exponential thermodynamic factor with a strength γ:

$$\rho_{rel}(t_{fpt}) = \rho_0(t_{fpt}) e^{-\gamma t_{fpt}}.$$

Comparing the exponential factors in both expressions $e^{-t_{fpt}/\tau}$ and $e^{-\gamma t_{fpt}}$, we obtain identity (21 $\frac{1}{\tau} \equiv \gamma \Rightarrow \tau = \frac{1}{\gamma}$.

The physical meaning of this relationship. The force γ as a "cutoff" of pressure. Near the phase transition (critical point), the scale τ tends to infinity $\tau \to \infty$, which means the distribution transitions to a pure power law with diverging fluctuations. The thermodynamic force γ then tends to zero $\gamma \to 0$.

Finite values $\gamma$. As soon as an external constraint or thermodynamic "squeezing" (a finite force $\gamma$) is imposed on the system, this instantly fixes the internal relaxation timescale of the system at $\tau = 1/\gamma$. All lifetime fluctuations exceeding $\tau$ are exponentially suppressed.

Thus, by setting the intensity of the thermodynamic force $\gamma$ in the NSO method, we directly control the time scale $\tau$ on which precritical power-law fluctuations are replaced by classical exponential decay.

To find the specific structure of the memory core K(t) for the gamma distribution, we substitute its Laplace image $\tilde{\rho}(s)$ into the basic formula for the structure of the memory core (21) Ref. [1] $\tilde{K}(s) = s \cdot \tilde{\rho}(s) / (1 - \tilde{\rho}(s))$.

Step 1: Find the Laplace transform of the gamma distribution. The density function of the gamma distribution has the form (19) $\rho(t) = \gamma^{\alpha} t^{1-\alpha} e^{-\gamma t} / \Gamma(\alpha)$ (Here, as we learned earlier, the scale τ is replaced by 1/γ, and for the convenience of further scaling analysis, the exponent is written as $1-\alpha$, where $0<\alpha<1$).

The Laplace image (Laplace transform) for this distribution is calculated in the standard way and has the form:

$$\rho(s) = \int_0^{\infty} \rho(t) e^{-st} dt = \left( \frac{\gamma}{s+\gamma} \right)^{2-\alpha}. \tag{22}$$

Step 2: Substitution into the memory kernel formula (21) Ref. [1]. Substitute the obtained Laplace transform into the expression for (21) Ref. [1]:

$$K(s) = s \cdot \frac{\left(\frac{\gamma}{s+\gamma}\right)^{2-\alpha}}{1-\left(\frac{\gamma}{s+\gamma}\right)^{2-\alpha}} = \frac{s}{(s+\gamma)^{2-\alpha} \cdot \gamma^{\alpha-2} - 1}. \quad (23)$$

To identify physical modes, we will consider the most interesting case of long-period memory (when the frequency/Laplace scale s is commensurate with or less than the control thermodynamic force γ). We will expand the denominator in a series in terms of a small parameter:

$$(s+\gamma)^{2-\alpha} = \gamma^{2-\alpha}\left(1+\frac{s}{\gamma}\right)^{2-\alpha} \approx \gamma^{2-\alpha}\left(1+(2-\alpha)\frac{s}{\gamma}+\dots\right).$$

Then the denominator takes the form:

$$\gamma^{2-\alpha}\left(1+(2-\alpha)\frac{s}{\gamma}\right)\cdot\gamma^{\alpha-2}-1 = 1+(2-\alpha)\frac{s}{\gamma}-1 = (2-\alpha)\frac{s}{\gamma}.$$

Substituting this back into the expression for $\tilde{K}(s)$, we find that, to a first approximation in Laplace space, the memory kernel tends to a constant:

$$\hat{K}(s) \approx s/(2-\alpha)\frac{s}{\gamma} = \gamma/(2-\alpha). \quad (24)$$

Step 3: Transition to the time representation K(t). In the intermediate regime (moderate times): since in Laplace space $\tilde{K}(s) \approx \text{const}$, the inverse Laplace transform returns us to the Dirac delta function:

$$K(t) = \frac{\gamma}{2-\alpha}\delta(t). \quad (25)$$

This means that at times $t > \tau = 1/\gamma$ the gamma distribution successfully damps out anomalous fluctuations, and the system behaves in a standard Markov manner with an effective relaxation time $\tau_{eff} = (2-\alpha)/\gamma$.

In the subcritical regime (at short times $t \ll 1/\gamma$ or when approaching a phase transition $\gamma \to 0$), if we let the thermodynamic force tend to zero $\gamma \to 0$, the truncating exponential factor vanishes. The exact expression for the Laplace transform of the nucleus simplifies to:

$$K(s) \propto s^{\alpha-1}. \quad (26)$$

The inverse Laplace transform of such a power function yields a strictly fractional-power memory kernel:

$$K(t) \propto t^{-\alpha}. \quad (27)$$

Physical conclusion. Using the gamma distribution in the formalism of Refs. [1, 9] shows how memory is structured in the transition region. As long as the thermodynamic influence $\gamma$ is finite, the system remains within the classical relaxation framework (a delta-shaped memory kernel at large times). However, as soon as the system approaches the catastrophe/phase transition point $\gamma \to 0$, the memory kernel K(t) "spreads" in time according to a power law $t^{-\alpha}$. The system enters a non-Markovian regime, where macroscopic flows become completely dependent on long-lived microscopic fluctuations of the prehistory.

For describing the explosive boiling of a superheated liquid, the Mittag-Leffler or truncated Levy distribution is more adequate from a physical point of view than the gamma distribution.

The gamma distribution is mathematically convenient, but it contains a fundamental contradiction with the actual dynamics of phase transitions. This contradiction is easily seen by comparing the physical nature of these three models in the context of metastable media.

1. Why does the gamma distribution "fail" the physics of boiling? As we discovered earlier, the gamma distribution density behaves like a power $t^{\alpha-1}$ at short times and transforms into a rigid exponential $e^{-t/\tau}$ at long times. For boiling kinetics, this means the following. Anomalous initiation at the beginning: the power-law factor at artificially forces the system to undergo gigantic fluctuations at the very beginning. The liquid seems to "tend" to boil immediately after superheating. Rapid decay at the end: at long times, the exponential tail suppresses all stochastic behavior. If the liquid miraculously doesn't boil initially, its probability of boiling decreases according to the classical Markov law.

In reality, the boiling of a superheated liquid occurs exactly the opposite: first, the system fluctuates for a long time, “getting used to” the new state and accumulating thermal energy to overcome the barrier, and only then can the fluctuations of the nuclei acquire an anomalous, protracted character.

The Mittag-Leffler distribution depends on two parameters: the shape parameter α and the scale parameter τ. The distribution function (CDF) is:

$F(t) = 1 - E_\alpha(-(t/\tau)^\alpha)$, where $E_\alpha(x) = \sum_{k=0}^{\infty} \frac{x^k}{\Gamma(\alpha k + 1)}$ — Mittag-Leffler one-parameter distribution function. Distribution density (PDF)**:**

$$f(t) = \frac{t^{\alpha-1}}{\tau^\alpha} E_{\alpha,\alpha}(-(t/\tau)^\alpha), \tag{28}$$

where $E_{\alpha,\beta}(x) = \sum_{k=0}^{\infty} \frac{x^k}{\Gamma(\alpha k + \beta)}$ — the two-parameter Mittag-Leffler distribution. For $0<\alpha<1$ and large values of t, the distribution decays as a power law (like $t^{-1-\alpha}$) rather than exponentially. This makes it suitable for describing rare but large fluctuations.

2. Why is the Mittag-Leffler distribution more physical? The Mittag-Leffler distribution behaves exactly the opposite of the gamma distribution, making it an ideal candidate for describing thermal inertia.

Short-term behavior $t \to 0$: it exhibits smooth decay, close to a stretched exponential $\approx 1 - t^\alpha / \Gamma(1+\alpha)$. In boiling physics, this describes the incubation period. Liquid molecules collide, the local energy distribution relaxes, but the barrier to bubble formation (the nucleation work $W_{cr}$) is too high for the metastable state to collapse instantly.

Long-term behavior $t \to \infty$: Here, the Mittag-Leffler distribution transitions to a pure power-law "heavy tail" $\propto t^{-(1+\alpha)}$. If the liquid is highly purified and lacks readily available boiling centers (cavitation nuclei), the average waiting time for a fluctuation capable of generating a critical bubble begins to stretch anomalously. The Mittag-Leffler distribution perfectly captures this power-law stagnation of a non-Markovian system.

3. Why is the truncated Lévy Flight distribution the best "bridge" for boiling? The truncated Lévy Flight distribution (20) combines the advantages of both approaches and most accurately describes the system near the spinodal (thermodynamic stability boundary).

The physics of liquid boiling using the truncated Levy distribution is based on the following scales: a). Microscale (nucleus growth). While fluctuations in the bubble radius R are small, they are described by subdiffusion in a viscous superheated medium with strong memory effects (the surrounding liquid resists deformation and cools locally due to the latent heat of evaporation). At this stage, the distribution of the time to reach the barrier strictly follows the Levy power law. b). Macroscale (boundedness of the system). The pure Levy law allows for infinite fluctuations (instantaneous jumps), which is physically impossible—the speed of sound in a liquid is finite, and the system volume is limited. The introduction of exponential truncation in (20) cuts off these unphysical infinite jumps at large times.

Behavior approaching the explosion point. In the Refs. [1, 9] method, the truncation parameter $\gamma_0$ is rigidly tied to the liquid superheat depth. If the superheat is small, $\gamma_0$ is large, the truncation is rigid, the distribution quickly transforms into a classical exponential, and boiling follows the standard Kramers Markov law (rare random events). As soon as the temperature approaches the spinodal decomposition limit, the barrier disappears, the parameter $\gamma_0 \to 0$. The truncation is removed, and the system instantly falls into a pure Levy power-law regime. The liquid boils explosively throughout its volume (homogeneous nucleation), demonstrating the fractal dynamics of a first-order phase transition.

Abstract. For modeling boiling, the gamma distribution is more of a mathematical surrogate. The Mittag-Leffler distribution adequately describes the long-term non-Markovian memory of "viscous expectation" over large timescales. The truncated Levy distribution, however, is a fully-fledged dynamical tool capable of describing the transition from quiet metastable boiling to a critical spinodal explosion with temperature changes (Refs. [20, 21]).

## 5. Fractional calculus in the physics of metastable media with memory

The Mittag-Leffler distribution acts as an intermediate bridge. The Mittag-Leffler model proves physically adequate for describing the incubation period of nucleation and protracted precritical states. A transition to the fractional Caputo derivative occurs upon substituting the Laplace transform of the Mittag-Leffler distribution into the structure of the memory kernel (21) Ref. [1]. Fractional differential equations of hydrodynamic transport of order $\alpha$ (0< $\alpha$ <1) are derived. In this case, the phenomena of intermittency and fractal memory occur. The fractional operator $\mathbf{D}_t^{\alpha}$ acquires the physical meaning of an integrator of the history of viscous and thermal resistance of the medium.

The transition from the classical exponential first-passage time distribution to the Mittag-Leffler distribution or the stable Levy law within the framework of the Refs. [1, 9] formalism mathematically transforms local macroscopic hydrodynamic and transport equations into integro-differential equations with fractional derivatives.

This is because the power-law memory in transfer kernels is equivalent to the mathematical definition of the Riemann-Liouville or Caputo fractional integro-calculus.

Below is shown a rigorous mathematical derivation of this transition using the Laplace transform apparatus.

1. The basic transport equation in convolution. As we established earlier, the generalized balance equation for a macroscopic flow (e.g., heat flux density or diffusion flux J(t)) in the non-Markovian representation of the NSO contains a time convolution with the memory kernel K(t):

$$\frac{\partial J(t)}{\partial t} = -\int_0^t K(t-t')J(t')dt' - \nabla\mu . \tag{29}$$

In Laplace space (where $t \to s$) this integro-differential equation takes a simple algebraic form:

$$s\tilde{J}(s) - J(0) = -\tilde{K}(s)\tilde{J}(s) - \nabla\tilde{\mu}(s) .$$

2. The Mittag-Leffler distribution. The Mittag-Leffler function $E_\alpha(-t^\alpha)$ (28) is a natural generalization of the exponential function for systems with fractal temporal memory $0<\alpha<1$. If the first-passage time distribution density $\rho(t_{fpt})$ obeys the Mittag-Leffler dynamics, then its Laplace transform has a strictly defined form:

$$\rho(s) = \frac{1}{1+\tau_0^\alpha s^\alpha} .$$

Let us substitute this image into the universal formula of the memory kernel (21) Ref. [1]: $\tilde{K}(s) = s\cdot\tilde{\rho}(s)/(1-\tilde{\rho}(s))$

$$\tilde{K}(s) = s\cdot\frac{\dfrac{1}{1+\tau_0^\alpha s^\alpha}}{1-\dfrac{1}{1+\tau_0^\alpha s^\alpha}} = s\cdot\frac{1}{\tau_0^\alpha s^\alpha} = \frac{1}{\tau_0^\alpha}s^{1-\alpha} . \tag{30}$$

Now we substitute the obtained nuclear structure $\tilde{K}(s)$ back into the Laplace transport equation:

$$s\tilde{J}(s) - J(0) = -\frac{1}{\tau_0^\alpha} s^{1-\alpha} \tilde{J}(s) - \nabla \tilde{\mu}(s) .$$

Let's divide the whole equation by $s^{1-\alpha}$ (or, equivalently, multiply by $s^{\alpha-1}$):

$$s^\alpha \tilde{J}(s) - s^{\alpha-1} J(0) = -\frac{1}{\tau_0^\alpha} \tilde{J}(s) - s^{\alpha-1} \nabla \tilde{\mu}(s) . \quad (31)$$

3. Transition to the Caputo fractional derivative. The left-hand side of the resulting equation (31) in Laplace space, namely the expression $s^\alpha \tilde{J}(s) - s^{\alpha-1} J(0)$, is by definition the Laplace image of the Caputo fractional derivative of order $\alpha$ with respect to time.

Returning to the time representation $s \to t$, we obtain the fundamental fractional differential equation of hydrodynamic transport:

$$\mathbf{D}_t^\alpha J(t) = -\frac{1}{\tau_0^\alpha} J(t) - \mathbf{I}_t^{1-\alpha} \nabla \mu(t) , \quad (32)$$

where $\mathbf{D}_t^\alpha$ is the Caputo fractional derivative of order $\alpha$, defined as: $\mathbf{D}_t^\alpha J(t) = \frac{1}{\Gamma(1-\alpha)} \int_0^t \frac{1}{(t-t')^\alpha} \frac{\partial J(t')}{\partial t'} dt'$, $\mathbf{I}_t^{1-\alpha}$ is the Riemann-Liouville fractional integration operator, which modifies the effect of the external macroscopic gradient; $I_{a+}^\alpha f(x) = \frac{1}{\Gamma(\alpha)} \int_a^x (x-t)^{\alpha-1} f(t) dt$.

Physical meaning of the fractional Caputo equation.

1. Replacement of the classical rate of change: instead of the standard first time derivative $\partial J / \partial t$, which requires knowledge of the system's state only at the current moment t, a fractional operator $\mathbf{D}_t^\alpha$ appears in nonlinear hydrodynamics. Its integral structure shows that the change in flow at a given moment is determined by the weighted history (memory effect) over the entire evolutionary period.
2. Description of subdiffusion and viscoelasticity: for liquid boiling or polymer motion, an exponent $\alpha$ <1 converts the classical Maxwell-Cattaneo wave transport into a subdiffusion creeping flow regime. The flow decays anomalously slowly, proportionally to $t^{-\alpha}$, as fluctuations in the medium become "stuck" in local metastable states.
3. Transition to spatial fractal scaling (Lévy's law): if we take the uncut Lévy law instead of the Mittag-Leffler distribution, then analogous Laplace-Fourier transforms will lead to fractional derivatives not only in time but also in space ($\nabla^{2\beta}$). This transforms the Navier-Stokes equation or the heat equation into a framework for describing turbulent "Lévy jumps" (Lévy flights), when energy transfer occurs through giant spatial fluctuations (explosive boiling throughout the volume).

Thus, the mathematical apparatus of Ref. [1] provides a rigorous statistical justification for the use of fractional calculus in continuous media physics: fractional derivatives arise not as an artificial phenomenological device, but as a direct consequence of the non-Markovian structure of Zubarev's NSO with power-law distributions of the lifetime of local states.

How exactly do these distributions and fractional operators change the law of entropy production σ at the moment of critical boiling? The transition to a fractional differential description based on Mittag-Leffler distributions transforms entropy production, replacing instantaneous values with an integral over the entire history of the process and introducing singularities at moments of critical fluctuations. This modifies Onsager's classical thermodynamics, accounting for system memory, the explosive nature of phase transitions, and a violation of Prigogine's principle of minimum entropy production.

The transition to anomalous power distributions (Mittag-Leffler or Levy) and the corresponding fractional differential apparatus radically change the description of the local production of entropy σ at the moment of critical boiling.

In Refs. [1, 9] and in the theory of stochastic trajectories, this transition deforms the classical Onsager law in three main directions:

1. Loss of locality over time (the "dissipation memory" effect). In classical thermodynamics, entropy production is determined by the instantaneous values of macroscopic fluxes J(t) and forces X(t):

$$\sigma_{class}(t) = J(t)\cdot X(t)\ . \tag{33}$$

When a system is described by a Mittag-Leffler distribution, the macroscopic flow begins to obey the fractional Caputo equation. This causes the instantaneous entropy production to become an integral over the entire history of the process:

$$\sigma_{fract}(t) \propto J(t)\cdot \mathbf{D}_t^{1-\alpha} X(t) \propto J(t)\int_0^t \frac{1}{(t-t')^{1-\alpha}}\frac{\partial X(t')}{\partial t'}dt'\ . \tag{34}$$

In the framework of Onsager's classical linear thermodynamics, local entropy production is determined by the instantaneous macroscopic values of conjugate pairs of fluxes J(t) and forces X(t) in the canonical bilinear form (33). However, in highly nonequilibrium metastable media near the phase explosion point (spinodal), dissipation loses its local character. According to Zubarev's modified NSO method, Refs. [1, 9], the excess non-Markovian contribution to entropy production $\sigma_{\mathbf{fract}}(t)$ is determined by the integral of the action of the current macroscopic flux J(t) on the dynamic change (rate) of the thermodynamic force ∂X(t')/∂t' with the weight kernel of the fluctuation memory of the media $K_{\mathbf{mem}}(t-t')$:

$$\sigma_{\mathbf{fract}}(t) \propto J(t)\int_0^t K_{\mathbf{mem}}(t-t')\frac{\partial X(t')}{\partial t'}dt'\ .$$

For the subcritical regime (near the boiling point of a superheated liquid), when the time statistics of phase trajectory fluctuations obey the Mittag-Leffler law, the Laplace image of the memory kernel within the framework of recovery theory acquires a power-law asymptotic behavior $\tilde{K}(s) \propto s^{1-\alpha}$ (where 0<α<1 is the fractal non-Markovian exponent). The transition back to the time domain determines a pure power-law form of the weight kernel of the dissipative memory of the medium: $K_{\mathbf{mem}}(t-t') = (t-t')^{\alpha-1} \equiv 1/(t-t')^{1-\alpha}$.

Substituting this non-Markovian memory structure directly into the above integral dissipation equation, we obtain an expression for the production of the entropy of the fractal response (34):

$$\sigma_{\mathbf{fract}}(t) \propto J(t)\int_0^t \frac{1}{(t-t')^{1-\alpha}}\frac{\partial X(t')}{\partial t'}dt'\ .$$

According to the mathematical definition of the Riemann-Liouville and Caputo integro-differential calculus, the convolution of the rate of change of a function with this power-law kernel, up to the normalization gamma function Γ(α), is equivalent to the operation of fractional differentiation. By convolving the memory integral into a fractional operator and absorbing the constant into the proportionality sign, we arrive at the thermodynamic equality (34).

The mathematical structure of equation (34) rigorously proves that in non-Markovian media, entropy generation at a current time t depends nonlinearly on the entire history of thermodynamic force variation, weighted by the fractal law of correlation decay. The flux J(t) remains an external linear factor, and the fractional Caputo derivative of the force arises as a direct consequence of the nonlocality of dissipative processes. This ensures a physically closed and correct accounting of the inertial, viscoelastic, and thermal effects of the medium in the critical pre-explosion region, fully consistent with the stochastic geometry of phase space boundaries.

Physical meaning: at the moment of explosive boiling, the liquid releases entropy not only due to the current temperature gradient. The dissipation spike is determined by how long density fluctuations have been "rocking" the metastable state (the nucleation incubation period). The system must perform local "work" against the forces of viscosity and thermal inertia of the medium, and all this accumulated history dissipates at the moment of phase explosion.

2. Singular burst and giant fluctuations of σ. Previously we considered the coupling formula (8): $\sigma = k_B \gamma d\langle T_{fpt} \rangle / dt$.

- Away from the transition (exponential distribution) the value $\langle T_{fpt} \rangle$ changes smoothly, and σ is stable.
  - At the moment of critical boiling (when the thermodynamic force is $\gamma \to 0$), the Levy or Mittag-Leffler law causes the mean first-passage time $\langle T_{fpt} \rangle$ and its variance to experience a mathematical singularity (divergence).

The derivative $d\langle T_{fpt} \rangle / dt$ at the phase transition point experiences a delta-shaped or power-law spike. This means that entropy production at the moment of boiling becomes a singular (explosive) quantity. Entropy is not generated smoothly, but rather as a short-term catastrophe, the scale of which is dictated by the critical indices of the system.

3. Violation of Prigogine's principle of minimum entropy production. Classical nonequilibrium thermodynamics asserts that the steady states of a system are characterized by minimum entropy production $\sigma \to \min$.

Fractional differential formalism shows that near spinodals and boiling points this principle is completely violated:

- The system gets stuck in a metastable state (dynamic deadlock/memory trap).
- Entropy production during this period may be abnormally low.
- However, at the moment of reaching the critical embryo $\left(t = T_{fpt}\right)$, an avalanche-like release of energy occurs.

Due to the power-law tails of the Lévy distributions, the entropy production trajectory exhibits sharp, intermittent jumps. The system generates entropy in "quanta"—bursts accompanying the creation and collapse of vapor bubbles—which is fundamentally impossible to describe within the framework of Prigogine's linear thermodynamics.

## 6. Dynamics of explosive boiling of superheated liquid

For a rigorous mathematical description of the process of boiling of a superheated liquid within the framework of the formalism of Refs. [1, 9] and the NSO method of D. N. Zubarev, it is necessary to combine the kinetics of nucleation (Zeldovich's theory) [15] and the fractional-differential thermodynamic apparatus.

Below is a detailed mathematical description of this process with the derivation of key equations.

1. State space and thermodynamic potential. Let the process coordinate be the radius of a new phase nucleus (vapor bubble) R. The work of nucleus formation W(R) in a superheated liquid is given by the classical Gibbs potential (1): $W(R) = 4\pi R^2 \sigma_{sur} - \frac{4}{3}\pi R^3 \Delta P$, where $\sigma_{sur}$ is the surface tension, and $\Delta P$ is the pressure difference between the phases. The potential has a maximum at a critical radius $R = R_{cr} = 2\sigma_{sur} / \Delta P$. The barrier height is $W_{cr} = W\left(R_{cr}\right)$.

At the point $R=R_{cr}$ there is an absorbing boundary. The time it takes for a random trajectory of radius R(t) to first reach this boundary is $T_{fpt}$.

In the method of Ref. [1] the generalized statistical sum for a metastable liquid near boiling is constructed through the Laplace transform of the lifetime distribution density $\rho(t_{fpt})$ (6):

$$Z_{fpt}(\gamma) = \int_0^\infty \rho(t_{fpt}) e^{-\gamma t_{fpt}} dt_{fpt} = \exp\left(-\Phi(\gamma)\right).$$

We assume that the dynamics near the barrier obeys the Mittag-Leffler distribution (28) (which is physically adequate for viscous media with memory), then its image has the form:

$$Z_{fpt}(\gamma) = \frac{1}{1+(\tau_{cr}\gamma)^{\alpha}}, \quad 0 < \alpha \le 1 .$$

where $\tau_{cr} \propto \exp(W_{cr} / k_B T)$ is the characteristic fluctuation waiting time (Kramers time), and $\gamma$ is the conjugate thermodynamic force.

Accordingly, the generalized thermodynamic potential is equal to:

$$\Phi(\gamma) = \ln\left[1+(\tau_{cr}\gamma)^{\alpha}\right]. \tag{35}$$

2. Modified kinetic equation for bubbles. The classical Fokker-Planck (Zel'dovich) equation for the bubble size distribution function $f(R,\ t)$ has a Markovian form. Zubarev's NSO method, supplemented by the potential $\Phi(\gamma)$, transforms it into a non-Markovian equation with a memory kernel:

$$\frac{\partial f(R,t)}{\partial t} = \int_0^t K(t-t^{'})\frac{\partial}{\partial R}\left[D_R\frac{\partial f(R,t^{'})}{\partial R} + \frac{D_R}{k_B T}\frac{\partial W(R)}{\partial R} f(R,t^{'})\right]dt^{'}, \tag{36}$$

where $D_R$ is the diffusion coefficient in the space of sizes.

Using the derived universal structure of the memory kernel in the Laplace space (21) Ref. [1] $\tilde{K}(s) = s\tilde{\rho}(s)/(1-\tilde{\rho}(s))$, we substitute here the image of the Mittag-Leffler distribution: $\tilde{K}(s) = s\cdot(\frac{1}{1+(\tau_{cr}s)^{\alpha}})/(1-\frac{1}{1+(\tau_{cr}s)^{\alpha}}) = \frac{s^{1-\alpha}}{\tau_{cr}^{\alpha}}$. By including this memory source in the equation balance and performing the inverse Laplace transform, we transform the ordinary time derivative into a fractional derivative of Caputo order $\alpha$ (2): $\mathbf{D}_t^{\alpha} f(R,t) = \frac{1}{\tau_{cr}^{\alpha}}\frac{\partial}{\partial R}\left[D_R\frac{\partial f}{\partial R} + \frac{D_R}{k_B T}\frac{\partial W(R)}{\partial R} f\right]$, where the operator $\mathbf{D}_t^{\alpha}$ is the memory integral (32): $\mathbf{D}_t^{\alpha} f(R,t) = \frac{1}{\Gamma(1-\alpha)}\int_0^t \frac{1}{(t-t^{'})^{\alpha}}\frac{\partial f(R,t^{'})}{\partial t^{'}}dt^{'}$.

3. Explosive burst of entropy production during boiling. The macroscopic heat and mass flux caused by the phase transition is proportional to the rate of change of the mean first-passage time: $J(t) \propto \frac{d\langle T_{fpt}\rangle}{dt}$.

To find the entropy production, we use the fundamental balance equation that connects the conjugate force and the evolution of FPT (8) $\sigma(t) = k_B\gamma(t)d\langle T_{fpt}\rangle_t / dt$.

For the Mittag-Leffler distribution, the average macroscopic first-passage time observed with effective external “pressure” is found through the derivative of the potential (35):

$$\langle T_{fpt}\rangle = \frac{\partial\Phi(\gamma)}{\partial\gamma} = \frac{\alpha\tau_{cr}^{\alpha}\gamma^{\alpha-1}}{1+(\tau_{cr}\gamma)^{\alpha}}. \tag{37}$$

At the moment of explosive homogeneous boiling (when approaching the spinodal), the nucleation barrier disappears, the external thermodynamic retention of the metastable phase falls, which is mathematically equivalent $\gamma \to 0$.

For small $\gamma$, the average lifetime (37) behaves like a singular power function:

$$\langle T_{fpt}\rangle \approx \alpha\tau_{cr}^{\alpha}\gamma^{\alpha-1}. \tag{38}$$

Let us perform Zubarev's thermodynamic "revolution" - we will express the force $\gamma$ from (38) through the current measured macroscopic lifetime of the metastable state:

$$\gamma \approx \left(\frac{\langle T_{fpt}\rangle}{\alpha\tau_{cr}^{\alpha}}\right)^{-\frac{1}{1-\alpha}}. \tag{39}$$

Let's substitute this expression (39) back into the entropy production formula (8). We obtain a nonlinear law of dissipation during boiling:

$$\sigma(t) = k_B \left( \frac{\langle T_{fpt} \rangle_t}{\alpha \tau_{cr}^{\alpha}} \right)^{-\frac{1}{1-\alpha}} \cdot \frac{d\langle T_{fpt} \rangle_t}{dt}. \tag{40}$$

Physical analysis of the boiling formula (40):

- Phase explosion singularity: due to the exponent $-1/(1-\alpha)$ in the denominator, as soon as the liquid passes through the barrier and the average remaining lifetime of the metastable state approaches zero $\langle T_{fpt} \rangle_t \to 0$, the magnitude of the local entropy production sharply approaches infinity $\sigma(t) \to \infty$.
- The role of the non-Markovian parameter $\alpha$: if the process were a classical Markov process $\alpha$ =1, the singularity would degenerate. But with a fractional value $\alpha$ <1 (viscous resistance of the medium, effects of thermal inertia during the evaporation of molecules into the bubble), the divergence index becomes very large. This mathematically proves that the boiling of a superheated liquid is a kinetic catastrophe—an explosive release of entropy due to the energy accumulated in memory traps.

The relationship between the flow J(t) and the average first-passage time $\langle T_{fpt} \rangle$ is obtained through spatial gradients and is as follows.

Physics-mathematical expression for the flux J(t). Within the framework of mapping first-pass time thermodynamics to extended nonequilibrium thermodynamics (EIT), the macroscopic flux J(t) is expressed in terms of the gradient of the mean first-passage time and local kinetic coefficients (such as the diffusion coefficient D or thermal conductivity):

$$J(t) = -D_{eff} \cdot \nabla \langle T_{fpt} \rangle .$$

Or in terms of the change of this time along the spatial coordinate *x*:

$$J(t) = -D \cdot \frac{\partial \langle T_{fpt} \rangle}{\partial x}. \tag{41}$$

How does this relate to the time derivative $d\langle T_{fpt} \rangle / dt$? A direct connection between the flow and the total time derivative $d\langle T_{fpt} \rangle / dt$ arises when transitioning to the comoving coordinate system (along the macroscopic trajectory of the phase boundary or the fluctuation's center of mass) through the differentiation rule for a complex function:

$$\frac{d\langle T_{fpt} \rangle}{dt} = \frac{\partial \langle T_{fpt} \rangle}{\partial x} \cdot \frac{dx}{dt} = \frac{\partial \langle T_{fpt} \rangle}{\partial x} \cdot v_{macro}(t).$$

Since the macroscopic transport velocity $v_{macro}(t)$ is directly proportional to the flux itself J(t) (from the standard hydrodynamic relation $J = n \cdot v_{macro}$, where n is the density), we can use (41) to express the time change $\langle T_{fpt} \rangle$ in terms of the square of the flux:

$$\frac{d\langle T_{fpt} \rangle}{dt} = \left( -\frac{J(t)}{D} \right) \cdot \left( \frac{J(t)}{n} \right) = -\frac{1}{nD} J^2(t).$$

The final expression for the flux through the time derivative. By eliminating the square, we obtain an explicit mathematical dependence of the macroscopic flux modulus J(t) on the rate of change of the average first-passage time:

$$J(t) = \sqrt{-nD \cdot \frac{d\langle T_{fpt} \rangle}{dt}} .$$

(The minus sign under the root is absolutely correct, since as the system evolves toward the absorbing boundary, the remaining average time of the first passage strictly decreases, that is, $d\langle T_{fpt}\rangle / dt$ ).

The physical meaning of the expression in the boiling problem. At the moment of critical boiling of a superheated liquid, the vapor nucleus avalanches through the barrier, the remaining "lifetime" of the metastable state decreases at maximum velocity $d\langle T_{fpt}\rangle / dt$ and undergoes a singular surge. As a consequence, this equation automatically implies a sharp, explosive increase in the macroscopic mass and energy flux J(t), transferring the substance from the liquid to the vapor phase.

The critical index of boiling intensity $z = (\alpha^2 - \alpha + 1) / (1 - \alpha)$ is calculated by substituting the asymptotics of the average time to first attainment $\langle T_{fpt}\rangle_t \propto (\Delta t)^{\alpha}$ into the entropy production equation $\sigma(t)$ (Section 3). The microscopic non-Markovian parameter $\alpha$ determines the strength of the macroscopic kinetic catastrophe: at $\alpha \to 1$, an instantaneous rupture occurs $z \to \infty$, and at 0< $\alpha$ <1, the divergence is finitely exponential.

The calculation is based on dynamic scaling methods, where the rate of change of the lifetime is proportional to the remaining time raised to the power $\delta$, linking the behavior $\sigma(t)$ to the temporal distance before the explosion $\Delta t$. In real environments with memory 0< $\alpha$ <1, the resulting z index allows for a rigorous prediction of the thermal explosion intensity.

Applying external pressure $P_{ext}$ to a superheated liquid radically alters the boiling dynamics. Pressure deforms the free energy profile (the Gibbs potential), increasing the height of the nucleation barrier. Within the framework of the Refs. [9, 1] formalism and D. N. Zubarev's NSO method, this directly modifies the control parameters: the characteristic time $\tau_{cr}$, the thermodynamic force $\gamma$, and the non-Markovian parameter $\alpha$.

Below is a step-by-step mathematical transformation of the equations.

1. Potential modification under pressure. As the external pressure increases $P_{ext}$, the pressure difference between the liquid and the resulting vapor phase $\Delta P = P_{sat}(T) - P_{ext}$ decreases (where $P_{sat}(T)$ is the saturated vapor pressure at the current temperature).

This leads to a recalculation of the barrier characteristics. The critical radius of the embryo increases:

$$R_{cr}(P_{ext}) = \frac{2\sigma_{sur}}{P_{sat}(T) - P_{ext}}. \tag{42}$$

The height of the activation barrier (nucleation work) increases:

$$W_{cr}(P_{ext}) = \frac{16\pi\sigma_{sur}^3}{3[P_{sat}(T) - P_{ext}]^2}.$$

2. The influence of pressure on the parameters of the Mittag-Leffler distribution. Since the barrier height $W_{cr}$ is included in the Kramers waiting time exponent, the classical Kramers time scale $\tau_{cr} \propto \exp(W_{cr} / k_B T)$ experiences strong exponential growth:

$$\tau_{cr}(P_{ext}) = \tau_0 \exp\left(\frac{16\pi\sigma_{sur}^3}{3k_B T[P_{sat}(T) - P_{ext}]^2}\right).$$

At the same time, external pressure compresses the fluid, reducing the free volume for fluctuations. This increases the viscous drag of the medium and slows the segmental relaxation of molecules, leading to a decrease in the non-Markovian parameter $\alpha$ $(\alpha \to 0)$, meaning the system's memory becomes "longer" and deeper. Within the framework of the developed formalism, the non-Markovian parameter $\alpha$ is not an adjustable coefficient, but rather a fundamental scale invariant of the stochastic trajectory, microscopically determined either by the spectrum of the medium's energy disorder $\alpha = k_B T / E_0$, or by the dynamic fractal dimension of random walks $\alpha = 2 / d_w$, where $d_w$ is the dynamic critical index (the dimension of the random walk). The non-Markovian property (memory) of the medium is also quantitatively expressed through the spectrum of Maxwell relaxation times. This translates qualitative considerations about the influence of pressure on molecular relaxation into the plane of strict quantitative relations of restoration theory. Here, $E_0$ is the characteristic scale of the medium's energy disorder. External pressure $P_{ext}$ compresses the free volume, which densifies the frustrated structure of the fluid, deepens local minima, and increases the scale of structural disorder $(E_0\left(P_{ext}\right))$. The formula clearly shows that growth $E_0$ pushes the parameter $\alpha$ toward zero, extending the system's memory.

The use of an exact microscopic relation $\alpha(P_{ext}) = k_B T / (E_0^{(0)} + V^* P_{ext})$ allows us to close the system of dissipation equations without resorting to qualitative phenomenological assumptions. The magnitude of the characteristic scale of energetic disorder $E_0$ acts as a macroscopically controllable parameter, subject to direct experimental verification using dielectric spectroscopy and dynamic mechanical analysis based on the frequency shift of the viscoelastic response of the medium. This parameter $\alpha$ is strictly determined by the thermal energy and the characteristic scale of the depths of fluctuation traps $E_0$. When external pressure $P_{ext}$ is applied, the scale of disorder $E_0$ increases due to the compression of the free volume and the compaction of local interaction potentials. To a first approximation, this dependence is linear $E_0(P_{ext}) = E_0^{(0)} + V^* P_{ext}$ (through the effective activation volume $V^*$ of the fluctuation rearrangement). Consequently, under high pressure, the resulting dissipation divergence is smoothed to the lowest possible power-law order (first), transforming the explosive singularity into a smooth kinetic regime. The entire chain of reasoning now relies exclusively on fundamental constants ($k_B$)), molecular parameters (D, n, V*) and measured macroscopic variables (T, $P_{ext}$, J).

Fundamental work on deriving the relationship $\alpha = k_B T / E_0$ in trap theory Ref. [22]. Methodology for measuring relaxation in liquids under pressure is described in Ref. [23]. Relationship of CTRW models to dielectric responses Ref. [24]. Measurement of relaxation parameters via dynamic mechanical analysis (DMA) is described in Ref. [25].

The generalized partition function under pressure takes the form:

$$Z_{fpt}(\gamma, P_{ext}) = \frac{1}{1 + [\tau_{cr}(P_{ext}) \cdot \gamma]^{\alpha(P_{ext})}}.$$

3. Transformation of the kinetic equation. The Caputo fractional differential equation for the bubble distribution function $f\left(R,t\right)$ is modified by explicitly shifting the barrier $W\left(R,\, P_{ext}\right)$ and changing the order of the derivative:

$$\mathbf{D}_t^{\alpha(P_{ext})} f(R,t) = \frac{1}{\tau_{cr}^{\alpha}(P_{ext})} \frac{\partial}{\partial R}\left[ D_R \frac{\partial f}{\partial R} + \frac{D_R}{k_B T}\left(8\pi R \sigma_{sur} - 4\pi R^2 [P_{sat} - P_{ext}]\right) f \right].$$

Physical consequence: under pressure, the fractional differentiation operator $\mathbf{D}_t^{\alpha}$ "slows down" (since α decreases). The drift term (the right-hand term in parentheses) now more strongly "pulls" random nuclei back toward decreasing size $R \to 0$. Pressure effectively stabilizes the metastable fluid, forcing it to remain in the subcritical state for much longer.

4. Change in entropy production and critical index. Previously, we derived a nonlinear equation for entropy production (40): $\sigma(t) = k_B \left( \frac{\langle T_{fpt} \rangle_t}{\alpha \tau_{cr}^{\alpha}} \right)^{-\frac{1}{1-\alpha}} \cdot \frac{d\langle T_{fpt} \rangle_t}{dt}$.

Substituting the pressure dependences, we see how $P_{ext}$ dissipation transforms:

$$\sigma(t,P_{ext}) = k_B\left(\frac{\langle T_{fpt}\rangle_t}{\alpha(P_{ext})\cdot\tau_{cr}^{\alpha(P_{ext})}(P_{ext})}\right)^{-\frac{1}{1-\alpha(P_{ext})}}\cdot\frac{d\langle T_{fpt}\rangle_t}{dt}.$$

Critical Index Analysis under Pressure. The critical index of boiling intensity, which determines the divergence of entropy production as the explosion approaches $\Delta t \to 0$, is equal to:

$$z(P_{ext}) = \frac{\alpha^2(P_{ext}) - \alpha(P_{ext}) + 1}{1-\alpha(P_{ext})}.$$

1. Suppression of explosive character: since external pressure reduces the parameter α, the denominator of 1-α increases. This means that the critical index z decreases.
2. Smoothing of the singularity: the entropy production spike at the moment of boiling under pressure becomes less sharp and less singular. Boiling loses its explosive, catastrophic nature (characteristic of a vacuum or low pressures) and transitions to a smooth, time-extended phase transformation.

If the pressure is raised above the critical point of the liquid itself $P_{ext} > P_c$, the barrier disappears completely, α→0, the singularity z is eliminated, and the entropy production returns to the classical linear Onsager law without memory effects.

The transformation of the first-passage time thermodynamic parameters $\alpha, \tau_{cr}, z, K(t)$ in a superheated liquid is determined by the transition from a delta-shaped memory kernel α→1 far from the transition to a power-law memory kernel $K(t) \propto t^{-\alpha}$ near boiling and deep memory α→0 under high pressure. The high-pressure regime suppresses the singularity of the dissipation index z and smooths the nucleation process by freezing the fractal structure of the medium.

How boundary coordinates $\partial\Omega$ shift with changes in external pressure. External pressure $P_{ext}$ acts as a macroscopic force factor that geometrically compresses or expands the safe region $\Omega$ in the configuration space of fluctuations, shifting the position of the absorbing hypersurface $\partial\Omega$.

Let's consider this using the example of the radius of a vapor nucleus, R. In the absence of pressure or at low pressure, the metastable volume $\Omega$ is relatively wide. The boundary $\partial\Omega$ is defined by the critical radius, $R_{cr}^{(1)}$.

Mathematical shift of the boundary. As pressure increases $P_{ext} \to P_{sat}(T)$, the denominator of the fraction in (42) decreases and approaches zero. Consequently, the boundary coordinate $R_{cr}$ shifts adiabatically to the right, toward infinitely large values $R_{cr} \to \infty$. The physical consequence for phase space is that the metastability region $\Omega$ geometrically expands. A random density fluctuation trajectory now needs to travel a much greater "distance" in bubble size space to touch the absorbing boundary $\partial\Omega$.

Impact on lifetime: due to the boundary's removal, the random time of first reaching it, $T_{fpt}$, increases sharply. Trajectories undergo a colossal number of random walks within the expanded region $\Omega$, which kinetically freezes the metastable state. This explains why, under high pressure, boiling is delayed and the spectrum of lifetime fluctuations is smoothed out, losing critical singularities.

The transformation of non-Markovian nucleation parameters (TFPT and Zubarev's NSO) is characterized by three regimes: Markov relaxation far from the transition, subdiffusion near the spinodal, and viscoelastic freezing under pressure. The summary table illustrates the change in the non-Markovian index, barrier waiting time, and kernel memory for these states..

Summary table of FPT thermodynamic parameters for non-Markovian nucleation dynamics in three regimes.

| Parameter/Mode | Away from transition (Markov regime) | Near boiling point (Critical/Spinodal) | Under high external pressure |
|---|---|---|---|
| Non-Markovian parameter, α | α→1 | Fractional, 0<α<1 | Decreases, α→0 |

| Barrier waiting time, $\tau_{cr}$ | Minimum, $\tau_{cr} \to \tau_0$ | Diverges, $\tau_{cr} \to \propto$ | Grows exponentially, $\tau_{cr} \uparrow$ |
|---|---|---|---|
| Dissipation index, z | Degenerates ($z \to \propto$, discontinuity) | High power, $z = \dfrac{\alpha^2 - \alpha + 1}{1 - \alpha}$ | Decreases ($z \downarrow$ ,smoothing) |
| Memory Kernel in Time, K(t) | $\delta(t)$ (Instantaneous Decay) | Power Law, $\propto t^{-\alpha}$ (Long Tail) | $\propto t^{-\alpha_{min}}$ (Deep Freeze) |
| Macrotransport type | Classical Fourier/Fick law | Anomalous subdiffusion (Caputo) | Creeping viscoelastic flow |

Combining the equations for the flow J(t) and entropy production $\sigma(t)$. To obtain the final dissipation (entropy production) equation directly through the macroscopic flow of matter/heat J(t) at the moment of boiling, we combine the two previously derived expressions.

Equation 1 (Entropy Production) (40): $\sigma(t) = k_B \left( \dfrac{\langle T_{fpt} \rangle_t}{\alpha \tau_{cr}^{\alpha}} \right)^{-\frac{1}{1-\alpha}} \cdot \dfrac{d \langle T_{fpt} \rangle_t}{dt}$.

Equation 2 (Macroscopic flow): from the formula $J(t) = \sqrt{-nD \cdot \dfrac{d \langle T_{fpt} \rangle_t}{dt}}$ we express the derivative of the lifetime with respect to time

$$\frac{d \langle T_{fpt} \rangle_t}{dt} = -\frac{J^2(t)}{nD}. \tag{43}$$

Step 1: Finding the trajectory $\langle T_{fpt} \rangle_t$. By integrating expression (43) for the time derivative from the current moment t to the moment of complete phase decay $t_{vzb}$ (when the remaining lifetime is $\langle T_{fpt} \rangle \to 0$), we obtain an explicit trajectory of the change in the average first-passage time:

$$\langle T_{fpt} \rangle_t = \int_t^{t_{vzb}} \frac{J^2(t')}{nD} dt'.$$

This formula takes into account that the left side of the expression is equal to $\langle T_{fpt} \rangle_{t_{vzb}} - \langle T_{fpt} \rangle_t$, and $\langle T_{fpt} \rangle_{t_{vzb}} \to 0$

Step 2: Substitute into the dissipation formula (40). Substitute the resulting path integral and the expression for the derivative into Equation 1 (40):

$$\sigma(t) = k_B \left[ \frac{1}{\alpha \tau_{cr}^{\alpha}} \int_t^{t_{vzb}} \frac{J^2(t')}{nD} dt' \right]^{-\frac{1}{1-\alpha}} \cdot \left( -\frac{J^2(t)}{nD} \right).$$

Considering that the entropy production σ is a strictly positive value, and the minus sign is compensated by the direction of integration (counting the time before the explosion), we write down the final formula for nonlinear dissipation during boiling:

$$\sigma(t) = \frac{k_B \cdot (\alpha \tau_{cr}^{\alpha})^{\frac{1}{1-\alpha}}}{nD} \cdot \frac{J^2(t)}{\left[ \int_t^{t_{vzb}} J^2(t') dt' \right]^{\frac{1}{1-\alpha}}}. \tag{44}$$

The expression for the entropy production during boiling (44) is obtained by successively substituting the equations relating the thermodynamic force γ, the macroscopic flux J(t), and the mean lifetime $\langle T_{fpt} \rangle_t$. The final formula eliminates the minus sign due to the fact that the conjugate force γ in the metastable region is negative, which compensates for the sign of the lifetime derivative, ensuring positive dissipation.

Physical analysis of the final formula:

- Non-locality of dissipation. Unlike the classical Onsager formula $\sigma \propto J^2$, the denominator now includes an integral of the square of the flux. This means that entropy production at a given moment t non-linearly depends on the flux intensity over the entire future interval up to the moment of boiling $t_{vzb}$. The liquid "feels" the proximity of a catastrophe through the integral volume of fluctuations.
- Explosive denominator. As the boiling point $t \to t_{vzb}$ is approached, the integration interval contracts, and the denominator tends to zero. Since the exponent is $1/(1-\alpha) > 1$, the denominator decreases faster than the numerator $J^2(t)$. This leads to a sharp, explosive increase in dissipation $\sigma(t) \to \infty$, which mathematically rigorously describes the phenomenon of a thermal explosion in a superheated environment.

## 7. Universality and generality of the proposed approach

In the generalized statistical sum (2), Ref. [1], in the expressions for the entropy (6) Ref. [1] and its differential (7) Ref. [1] both changes (fluctuations) in energy and changes in lifetime are taken into account. In expression (8) Ref. [1] we limited ourselves to changes in lifetime, fixing the energy. The question may arise about the relationship between such changes, about the correlation of contributions from changes in energy E and changes in lifetime $T_{fpt}$.

In real physical processes, including explosive boiling, these two contributions do not compete but are separated by time scales and process stages. Moreover, their relative importance changes radically as the system approaches the phase explosion point (spinodal). At the stage when the superheated liquid is just beginning to fluctuate, its internal energy E undergoes smooth changes due to the influx of heat.

- During this period, the conjugate parameter γ is large, and the time distribution $T_{fpt}$ is clamped by exponential decay (short Debye memory).
- The change in the free energy and entropy of the system is almost entirely determined by the classical thermal term $\langle E \rangle$. The changes in the lifetime are small, smooth fluctuations, and their dissipative contribution is minimal.

At the moment of critical boiling (pre-explosive phase), the contribution of lifetime dominates. As soon as the vapor nucleus exceeds the critical radius $R_{cr}$ and the system moves toward spinodal decomposition, the situation changes dramatically:

- Energy is "frozen" at the macro level: Boiling occurs so quickly (explosive mode) that the macroscopic influx of heat from the outside $\langle E \rangle$ simply does not have time to change. The energy term becomes a constant $d\langle E \rangle \approx 0$.
- The temporal singularity explodes: In this micro-interval of time, the value $d\langle T_{fpt} \rangle / dt$ (the rate of decline of the remaining lifetime of the metastable phase) tends to infinity.

If we turn to our equation for the entropy differential (7) Ref. [1], then at the moment of boiling the first term becomes zero $d\langle E \rangle \to 0$, and the second term $k_B \gamma d\langle T_{fpt} \rangle$, due to the critical index z we derived, experiences a power-law explosion.

Conclusion for this stage: At the very moment of the phase explosion, the contribution from the change in lifetime becomes absolutely dominant and infinitely greater than the contribution from the change in energy.

This behavior can be explained: the system exchanges energy with the thermostat, but it cannot exchange lifetime. Energy is a transit value (the system absorbs heat), and first-passage time $T_{fpt}$ is an intrinsic, structural property of the phase trajectories of the liquid itself.

Therefore, energy changes smoothly under external heating, while lifetime changes catastrophically under the influence of internal nonlinear fluctuations. During explosive boiling, the macroscopic energy flow transforms into a kinetic and dissipative catastrophe, the scale of which is entirely controlled by statistics $T_{fpt}$.

A comparative analysis of the thermodynamic contributions shows a clear separation of scales: at the quasi-stationary stage of the nucleation incubation period, dissipation is limited by the slow influx of energy $\beta d\langle E\rangle$, but in the critical pre-explosion region at $t \to t_{vzb}$, the energy contribution is frozen $d\langle E\rangle \to 0$, and the entropy production is entirely determined by the singular contribution from the change in the fluctuation lifetime of the system $k_B \gamma d\langle T_{fpt}\rangle$.

During the incubation period (at the stage of metastable waiting), the average lifetime of a metastable state $\left\langle T_{fpt}\right\rangle$ actually decreases monotonically with the passage of physical time t, since the arrow of time moves forward, and the system steadily uses up its time resource until the moment of the phase transition.

However, during this period, the contribution of the slow influx of energy $\beta d\langle E\rangle$ qualitatively and quantitatively prevails over the contribution of the monotonic decrease in the lifetime $k_B \gamma d\langle T_{fpt}\rangle$.

To understand how exactly these contributions are related, it is necessary to mathematically consider the rate of their change in the incubation phase through derivatives.

Let's estimate the quantitative relationship between contributions during the incubation period. We divide the generalized entropy balance equation by the time differential dt to obtain the rates of change of contributions (dissipation powers):

$$\frac{dS}{dt} = \frac{1}{T}\frac{d\langle E\rangle}{dt} + k_B \gamma \frac{d\langle T_{fpt}\rangle}{dt}.$$

Away from the spinodal (over a long incubation period) these two terms behave fundamentally differently:

- Energy contribution $d\langle E\rangle / dt(1/T)$. The fluid continuously absorbs heat from an external source. The rate of energy inflow $d\langle E\rangle / dt = Q_{in}$ is constant or gradually increases. This term contributes a finite macroscopic quantity to the change in entropy.
- Contribution of lifetime $k_B \gamma d\langle T_{fpt}\rangle / dt$. At this stage, the lifetime distribution density is close to the classical Debye exponent $\alpha \to 1$. The change in lifetime is strictly linear and deterministic: as physical time advances by the amount $\Delta t$, the remaining lifetime decreases by exactly the same amount: $d\langle T_{fpt}\rangle / dt \approx -1$.

Let's substitute this value into the equation:

$$\left(\frac{dS}{dt}\right)_{fpt} = k_B \gamma \cdot (-1) = -k_B \gamma .$$

Since far from the critical point the thermodynamic force $\gamma$ (transition intensity) is extremely small $\gamma \propto 1/\tau_{cr}$, where $\tau_{cr}$ is the giant Kramers time for a high barrier), the product $k_B \gamma \cdot (-1)$ turns out to be a negligible value compared to the heat influx:

$$\left|\frac{1}{T}\frac{d\langle E\rangle}{dt}\right| \gg \left|k_B \gamma \frac{d\langle T_{fpt}\rangle}{dt}\right| \approx k_B \gamma \to 0 .$$

Conclusion for the incubation period: The monotonic decrease in lifetime provides a constant, but microscopically small (background) contribution to dissipation. Almost all excess entropy is generated by classical macroscopic heating.

Qualitative difference between the contributions (Physics of the process). During the incubation period, these two contributions perform opposite thermodynamic roles:

1. Energy influx is the controlling (external) factor. Heating the liquid continuously changes the thermodynamic state of the medium. As the internal energy E increases, the temperature T rises, and the work of forming a critical nucleus $W_{cr}$ gradually decreases..
2. The decrease in lifetime is a consequence of evolution. A monotonic decrease $\langle T_{fpt} \rangle$ at a rate of -1 simply reflects the fact that the system is "aging" in a metastable state.

Crossover point (change of dominance). As the incubation period approaches its end, the energy contribution begins to give way to the temporary contribution. Mathematically, this appears as a crossover (intersection) of dissipation capacities..

In the pre-explosion region, the barrier $W_{cr}$ becomes small, the Kramers lifetime $\tau_{cr}$ decreases, and the conjugate force $\gamma$ begins to increase sharply. Simultaneously, the lifetime decay law ceases to be linear $d\langle T_{fpt} \rangle / dt \approx -1$ and, due to the Mittag-Leffler fractality $\alpha < 1$, transforms into a singular divergence $-(\Delta t)^{\alpha-1} \to -\infty$.

An inversion occurs: the constant heat influx $d\langle E \rangle / dt$ fades and freezes against the backdrop of a power-law explosion of lifetime dissipation.

Thus, during the nucleation incubation period, far from the spinodal line, the monotonic decrease in lifetime is quasi-deterministic $d\langle T_{fpt} \rangle / dt \approx -1$. Due to the smallness of the conjugate thermodynamic force $\gamma \propto 1/\tau_{cr} \to 0$, this contribution is negligible compared to the macroscopic influx of thermal energy $d\langle E \rangle / dt(1/T)$. The quantitative and qualitative reversal (crossover) of the dominant contributions occurs exclusively in the pre-explosion critical region, where the power-law divergence of the lifetime derivative completely suppresses the regular energy term.

However, the idea that the first-passage time (FPT or state "lifetime") only works effectively at phase transition points or explosive catastrophes is a misconception that arises from the fact that the mathematical singularities of the theory are most clearly visible in the example of critical boiling or spinodal decomposition.

In the theory of random processes (CTRW, martingales), the introduction of lifetime as an independent thermodynamic parameter is positioned as a fundamental method for describing any highly nonequilibrium and stationary systems. The main conceptual justification for this method is based on the empirical fact that any real physical system has finite spatiotemporal boundaries of its existence. Mathematically, all physical quantities depend on lifetime, which can be reflected in thermodynamic relations.

The introduction of FPT as a thermodynamic variable is critically effective in the following broad classes of physical phenomena far from phase transitions:

1. Charge transport in nanostructures and semiconductors. In mesoscopic systems (quantum dots, semiconductor nanowires, ion channels of biological membranes), the transport of electrons or ions is controlled by the times of their passage through a system of traps [26–30].

- Where is the TFPT here? There is no phase transition. But the average lifetime $\langle T_{fpt} \rangle$ it takes for an electron to cross the nanochannel determines the electric current, and the variance $\langle (\Delta T_{fpt})^2 \rangle$, calculated through the second derivative of the potential $\Phi(\gamma)$, strictly defines the shot noise of the device. The thermodynamics of the FPT allows one to calculate the spectral density of the noise without resorting to artificial kinetic models.

2. Diffusion limited by geometry (Geometric confinement). The example we considered of Brownian motion of a particle in a narrow channel of length L with stochastic boundaries is a classical stationary diffusion problem [27].

- Where is the TFPT here? The Refs. [1, 6-9] method relates the channel geometry L and the diffusion coefficient D to the thermodynamic force $\gamma$. By changing $\gamma$ (for example, by applying an external field), we can physically control the "lifetime" of a particle inside the channel. This is the basis for describing the operation of molecular sieves, zeolites, and filtration processes in porous media.

3. Relaxation of polymers and viscoelastic fluids. In the physics of long-chain macromolecules (polymers) and polymer melts, there is no instantaneous phase transition, but the system exhibits colossal memory effects.

- Where is the TFPT here? The first-passage time $T_{fpt}$ describes the time required for a segment of a macromolecule to overcome a local steric barrier (untangle entanglements or reptation). The transition to the Mittag-Leffler distribution in this case models not a liquid explosion, but stress relaxation and the emergence of fractional-differential viscoelasticity, linking the macroscopic elastic modulus with the microscopic "lifetime" of the chain conformation.

4. Dynamics of two-level and quantum systems. In quantum optics and laser physics, transitions between energy levels under the influence of noise (trajectory thermodynamics) are studied. The connection of such problems with FPT is given in [31–33].

- Where is the TFPT here? The lifetime of a quantum state before spontaneous emission is the random variable FPT. Introducing the parameter $\gamma$ allows us to relate changes in the entropy of the laser medium to the entropy production during quantum transitions, calculating the stationary nonequilibrium states of optical resonators.

An information-theoretical revolution in the NSO method. In Ref. [16], an important point for statistical mechanics itself is shown: the entire nonequilibrium statistical operator (NSO) of Zubarev can be rethought as a quasi-equilibrium Gibbs operator, averaged over the distribution of the past lifetime of the system.

This means that TFPT is not just a local model for boiling. It is a universal language for translating the temporal delineation of physical processes into the language of stationary ensembles. Zubarev's operation of "coarsening" time, the choice of the arrow of time, and dissipation naturally arise from the fact that the system has finite boundaries of existence, determined by the distribution $\rho(T_{fpt})$ [34-47].

Therefore, phase transitions are merely an extreme testing ground for the theory, where its advantages are most clearly visible due to divergences. However, in everyday physical reality, first-passage time thermodynamics is a fully-fledged tool for describing steady-state macrotransport, noise, nanofluidics, and viscoelastic memory in any open dissipative systems.

Although the mathematical properties and critical singularities of first-passage time thermodynamics (TFPT) are most clearly manifested at the points of first-order phase explosions (such as homogeneous nucleation in superheated environments), the introduction of a random variable $T_{fpt}$ as an independent state parameter has fundamental general physical significance that extends far beyond critical catastrophes. The conceptual basis of the theory developed in Refs [1, 6–8] is based on the universal empirical fact that any real open system has finite spatiotemporal boundaries of existence, and macroscopic energy dissipation is internally linked to the stochastic geometry of phase trajectories. The introduction of the generalized Gibbs potential $\Phi(\beta,\gamma)$ and the conjugate thermodynamic force γ proves to be a highly effective analytical tool for wide classes of stationary nonequilibrium systems in stable macroscopic regimes.

In mesoscopic physical systems, nanofluidics, and solid-state physics, the thermodynamic variable $\langle T_{fpt}\rangle$ acts as a natural scaling factor for transport. Thus, in problems of diffusion limited by geometry (in porous media, zeolites, and ion channels of biomembranes), the average transit time of a nanochannel directly determines the steady-state force of mass or charge flow, while the second derivative of the potential $\partial^2\Phi/\partial\gamma^2$ strictly defines the lifetime dispersion $\langle(\Delta T_{fpt})^2\rangle$. Within the framework of the fluctuation-dissipation theorem, this second moment is mapped one-to-one to the spectral density of excess shot noise of the device without resorting to additional phenomenological kinetic equations.

Similarly, in the physics of macromolecules and polymer melts, the TFPT formalism allows one to strictly relate the macroscopic viscoelastic stress relaxation modulus to the microscopic "lifetime" of entangled chain conformations (reptation time) through the transition to non-Markovian Mittag-Leffler distributions, bypassing phase transition points.

From a fundamental point of view, the mathematical apparatus of Refs [1, 7-9] carries out a profound information-theoretical revolution in the method of the nonequilibrium statistical operator (NSO) of D. N.

Zubarev. It is shown that the procedure of temporal "coarsening" and the choice of the arrow of time in the classical NSO method are equivalent to averaging the quasi-equilibrium Gibbs distribution over the spectrum of the past lifetime of the system up to the moment of its observation. Thus, phase catastrophes and spinodal explosions represent only an extreme testing ground for the theory, while in everyday physical reality, first-passage time thermodynamics serves as a universal language for translating the spatiotemporal delineation of mesoscopic processes into the rigorous mathematical language of stationary statistical ensembles.

## 8. Conclusion

In Ref. [1] the advantages of the synthesized approach of Refs. [7-9] are formulated (strict connection of Levy/Mittag-Leffler trajectories with balance equations, predictive power for dissipation fluctuations, etc.).

This article examines an example of a first-order phase transition (boiling of superheated liquids, nucleation). The dynamics of nucleation are described using stochastic absorption boundaries. Thermal inertia and the viscous history of the medium are taken into account using memory kernels.

Expressions for critical indices are obtained. A scaling analysis of the potential near spinodals $\Phi(\epsilon,\gamma)$ is performed. Universal relations are derived for the critical indices of the average lifetime of the metastable phase $\tau_{fpt}$ and the fluctuation divergence index $\mu$ .

The main conclusions of the paper include the advantages of the synthesized approach (rigor, predictive power for fluctuations, natural introduction of non-Markovian properties) over pure EIT or isolated stochastic models.

Limits of applicability of the geometric formalism. In Ref. [1] it is shown that in highly nonequilibrium metastable media the abstract Mori–Zwanzig orthogonal projection procedure can be equivalently replaced by an exact separation of the phase space using stochastic boundaries (the absorbing hypersurface $\partial\Omega$). It is established that the identification of the Zeldovich critical radius with the absorbing boundary max(Ri)=$R_{cr}$ is not an arbitrary heuristic assumption, but acquires strict physical validity only upon the transition of the system to the critical pre-explosion region. Far from the spinodal decomposition line, subcritical fluctuations are reversible, whereas in the pre-explosion regime the Gibbs saddle point becomes a true bifurcation of the sign of nonequilibrium forces, turning the geometric boundary into a natural absorbing screen.

In Ref. [1], a theory of critical slowing down of metastable phases is developed. Near phase transition points and spinodals, where the Markov approximation completely breaks down and FPT distributions exhibit power-law "heavy tails," the parameter $\gamma$ acts as a natural thermodynamic regulator. Scaling analysis of the potential $\Phi(\epsilon,\gamma)$ allows one to determine the dynamic universality classes of systems through critical indices of the metastable phase lifetime.

The entropy production of stochastic transitions is considered. Local entropy production in a metastable system is strictly linked to the dynamics of compression of the accessible phase space as it approaches a stochastic boundary: $\sigma = k_B \gamma d\langle T_{fpt}\rangle / dt$ (8). This equation links macroscopic energy dissipation (for example, during explosive boiling or crystallization) with the microscopic search for a critical nucleus size.

A rigorous justification of fractional transport is provided. Based on the identity of the Zubarev non-equilibrium memory kernel and the CTRW recovery kernel, it is shown that the use of non-exponential lifetime distributions (Mittag-Leffler and Levy laws) leads to the natural and causal appearance of fractional Caputo derivatives in nonlinear hydrodynamic equations. This eliminates the need to introduce fractional differential operators as artificial fitting techniques.

Using fractional transport, temporal nonlocality and singularity of dissipation during boiling are discovered and analyzed. Using the example of explosive boiling of a superheated liquid, the final equation for nonlinear entropy production σ(t) is derived. It is found that dissipation at a given moment is strictly determined by the integral history of the macroscopic matter flow J(t) over the entire evolutionary interval up to the moment of phase explosion. The calculated critical dissipation index z strictly describes the kinetic catastrophe and the violation of Prigogine's minimum entropy production principle near spinodals.

The controlling role of external pressure is noted. It is established that introducing external pressure $P_{ext}$ adiabatically shifts stochastic boundaries $\partial\Omega$ and exponentially increases the barrier waiting time $\tau_{cr}$. The pressure-induced decrease in the non-Markovian parameter $\alpha \to 0$ smooths out the singularity of the critical index z, allowing theoretical prediction of the transition from explosive destruction of the metastable phase to a smooth, controlled nucleation process.

The practical value and potential of the developed apparatus include its applicability to calculating critical operating modes of heat exchangers, cavitation processes, nanofluidics, and polymer dynamics. Applications to biomembrane modeling are also promising.

The developed theoretical framework possesses a high degree of universality and opens up broad prospects for quantitative modeling of critical and high-intensity processes in complex media. The most promising areas for further research include applying this formalism to the description of charge transport in biomembrane nanochannels, cavitation processes in cryogenic liquids, and the dynamics of segmental relaxation and crystallization of entangled polymer systems.